\pdfoutput=1
\documentclass[11pt,twoside,a4paper,cmspaper,final,collab]{cms-tdr}
\def\svnVersion{99ccd45}\def\svnDate{2026/11/30}\def\cmsCernNoTag{CERN-EP-2026-072}\def\cmsCernDate{\today}\def\cmsMessage{Published in Physical Review Letters as \href{http://dx.doi.org/10.1103/vkyn-sp78}{\doi{10.1103/vkyn-sp78}.}}
\begin{document}\cmsNoteHeader{EXO-24-006}

\newlength\cmsFigWidth
\ifthenelse{\boolean{cms@external}}{\setlength\cmsFigWidth{0.49\textwidth}}{\setlength\cmsFigWidth{0.65\textwidth}}
\ifthenelse{\boolean{cms@external}}{\providecommand{\cmsLeft}{upper\xspace}}{\providecommand{\cmsLeft}{left\xspace}}
\ifthenelse{\boolean{cms@external}}{\providecommand{\cmsRight}{lower\xspace}}{\providecommand{\cmsRight}{right\xspace}}

\newcommand{\PY}{{\HepParticle{Y}{}{}}\Xspace}
\newcommand{\MX}{\ensuremath{m_{\PX}}\xspace}
\newcommand{\MY}{\ensuremath{m_{\PY}}\xspace}
\newcommand{\PeME}{\ensuremath{\Pe_{\mathrm{ME}}}\xspace}
\newcommand{\PmMM}{\ensuremath{\PGm_{\mathrm{MM}}}\xspace}
\newcommand{\vecClus}{\ensuremath{\vec{x}_{\text{clus}}}\xspace}
\newcommand{\vecEi}[1]{\ensuremath{\vec{x}_{\Pe_{#1}}}\xspace}
\newcommand{\vecEE}{\ensuremath{\vec{v}_{\Pe\Pe}}\xspace}
\newcommand{\ufxf}{\ensuremath{\mathrm{U}_{5\times5}}\xspace}
\newcommand{\ptratio}{\ensuremath{\abs{\sum{\vec{p}_{\mathrm{T}\Pell}}}/\ptmiss}\xspace}
\newcommand{\Mfl}{\ensuremath{m_{4\Pell}}\xspace}
\newcommand{\Mtl}{\ensuremath{m_{2\Pell}}\xspace}
\newcommand{\PellPass}{\ensuremath{\Pell_{\text{pass}}}\xspace}
\newcommand{\PellFail}{\ensuremath{\Pell_{\text{fail}}}\xspace}
\newcommand{\PeMEFail}{\ensuremath{\Pe_{\mathrm{ME}}^{\text{fail}}}\xspace}
\newcommand{\SD}{s.d.\xspace}

\newlength\cmsTabSkip\setlength{\cmsTabSkip}{1.3ex}

\ifthenelse{\boolean{cms@external}}{\renewcommand{\appendixname}{End Matter}}{}

\cmsNoteHeader{EXO-24-006}
\title{Search for heavy resonances decaying into four-lepton final states via light bosons in proton-proton collisions at \texorpdfstring{$\sqrt{s} = 13\TeV$}{sqrt(s) = 13 TeV}}

\date{\today}

\abstract{
  A search for a resonance heavier than 250\GeV decaying into four leptons via two intermediate bosons is presented. The search uses proton-proton collision data at $\sqrt{s} = 13\TeV$ collected by the CMS experiment, corresponding to an integrated luminosity of 138\fbinv. Novel techniques are used to enhance the sensitivity to a collimated pair of dileptons reconstructed as a single merged object, resulting from the decay of an intermediate light boson. No significant excess of data over the background predictions is observed. Upper limits are set on the production cross section for a four-lepton resonance, including the previously unexplored phase space at the LHC with a dilepton mass of 0.4--15\GeV.
}

\hypersetup{
pdfauthor={CMS Collaboration},
pdftitle={Search for heavy resonances decaying into four-lepton final states via light bosons in proton-proton collisions at sqrt{s} = 13 TeV},
pdfsubject={CMS},
pdfkeywords={merged electron, high-energy, boosted, four-lepton}}

\maketitle

Many theories beyond the standard model (BSM) predict the existence of a heavy boson that decays into two standard model (SM) or BSM bosons, which each subsequently decays into two-lepton (dilepton) final states. 
Models with an extended Higgs sector, such as the two-Higgs-doublet model (2HDM), typically predict a heavy neutral Higgs boson that decays into either a pair of pseudoscalar bosons, or a single pseudoscalar boson and a \PZ boson, where each particle decays into a dilepton pair~\cite{2HDMreview}. 
Four-lepton final states produced via intermediate bosons are also predicted by many theories with an additional $\mathcal{U}(1)$ gauge symmetry. These include models with dark or hidden sectors that propose a BSM particle that decays into lepton pairs via kinetic mixing with the SM photon and the \PZ boson~\cite{darkphoton}. Additionally, models motivated by grand unified theories suggest that a BSM heavy neutral boson may decay into four leptons if the direct coupling between a $\mathcal{U}(1)$ gauge boson and leptons is forbidden~\cite{ZprimePheno}.

Several studies have been performed to search for BSM physics in four-lepton final states at the CERN LHC. The most common searches are for rare decays of the Higgs boson into BSM particles~\cite{CMS_hZdZd,ATLAS_hXX} or heavy neutral particle decays into a pair of \PZ bosons~\cite{CMS_ZZ4l,ATLAS_ZZ4l}.
A search for a heavy scalar boson decaying into either a pair of new bosons or a new boson associated with a Z boson with the dilepton mass $>$15\GeV has also been reported~\cite{ATLAShZdZd}.
No sign of BSM physics has been observed within accessible mass ranges.

However, a light intermediate boson produced from a resonance significantly heavier than the SM Higgs boson can result in a high Lorentz boost of the dilepton system. This may cause the lepton pair to merge into a single reconstructed object or fail to be reconstructed at all, thereby potentially evading detection without specialized treatments.
We present a model-independent search for a high-mass resonance decaying into four leptons with the $\PX\to\PY\PY\to4\Pell$ and $\PX\to\PZ\PY\to4\Pell$ signal hypotheses ($\Pell = \Pe,~\PGm$). We probe $\MX$ in the range 250--2000\GeV and $\MY > 0.4\GeV$ within the kinematically allowed region, where \MX (\MY) represents the mass of the hypothetical boson \PX (\PY). 
The search uses proton-proton ($\Pp\Pp$) collision data at $\sqrt{s}=13\TeV$ collected by the CMS experiment in 2016--2018, corresponding to an integrated luminosity of 138\fbinv. Novel merged-electron (\PeME) and missing-muon (\PmMM) identification (ID) techniques are developed to enhance signal sensitivity.

The CMS apparatus~\cite{CMS:2008xjf,CMS:2023gfb} is a multipurpose, nearly hermetic detector, designed to trigger on~\cite{CMS:2016ngn,CMS:2020cmk,CMS:2024aqx} and identify electrons, muons, photons, and (charged and neutral) hadrons~\cite{CMS:2020uim,CMS:2018rym,CMS:2014pgm}. Its central feature is a superconducting solenoid of 6\unit{m} internal diameter, providing a magnetic field of 3.8\unit{T}. Within the solenoid volume are a silicon pixel and strip tracker, a lead tungstate crystal electromagnetic calorimeter (ECAL), and brass and scintillator hadron calorimeter (HCAL), each composed of a barrel and two endcap (EE) sections. The ECAL provides coverage in pseudorapidity $\abs{\eta} < 1.44$ in the barrel region (EB) and $1.57 < \abs{\eta} < 3.0$ in the two endcap regions. Muons are reconstructed using gas-ionization detectors interleaved with the layers of the steel flux-return yoke outside the solenoid.

A particle-flow (PF) algorithm~\cite{CMS:2017yfk} aims to reconstruct and identify each individual particle in an event, with an optimized combination of information from the various elements of the CMS detector. The energy of photons is obtained from the ECAL measurement. The energy of an electron is determined from a combination of the electron momentum at the primary interaction vertex as determined by the tracker and the supercluster (SC) energy, which incorporates the energy of the corresponding ECAL cluster and the energy sum of all bremsstrahlung photons spatially compatible with originating from the electron track.
The energy of a muon is obtained from the curvature of the corresponding track. The energy of a charged hadron is determined from a combination of the momentum measured in the tracker and the matching ECAL and HCAL energy deposits, corrected for the response function of the calorimeters to hadronic showers. Finally, the energy of neutral hadrons is obtained from the corresponding corrected ECAL and HCAL energies. The reconstructed particles are used to build \PGt leptons, jets, and missing transverse momentum (\ptmiss)~\cite{CMS:2018jrd,CMS:2016lmd,CMS:2019ctu}. 
The vector \ptvecmiss is computed as the negative vector sum of the transverse momenta (\pt) of all the PF candidates in an event, and its magnitude is denoted as \ptmiss~\cite{CMS:2019ctu}. The \ptvecmiss is modified to account for corrections to the energy scale of the reconstructed jets in the event~\cite{CMS:2020ebo}.

The benchmark signal events $\PX\to\PY\PY\to4\Pell$ and $\PX\to\PZ\PY\to4\Pell$ are generated using the \PYTHIA 8.240 Monte Carlo (MC) event generator~\cite{pythia8} for each $\MX = 250$, 500, 750, 1000, 1500, and 2000\GeV and $\MY = 0.4$, 0.6, 0.8, 1, 1.5, 2, 5, 10, 50, and 100\GeV. Signal events with $\MY = 250$, 500, 750, 1000, and 1500\GeV are also generated if kinematically allowed.
A narrow-width approximation is assumed for \MX and \MY by setting the width-to-mass ratio to $10^{-4}$, such that the detector dominates their reconstructed invariant mass distributions. 
The branching fractions $\mathcal{B}(\PY\to\Pe\Pe)$ and $\mathcal{B}(\PY\to\PGm\PGm)$ are each set to 0.5 for the combination of results involving different lepton flavors. 

The \ttbar, $\PW\PZ\to3\Pell\PGn$, and $\PZ\PZ\to4\Pell$ processes for the background estimation are also simulated using the \POWHEG v2.0~\cite{powheg1,powheg2,powheg3,Frixione:2007nw,powhegVV} event generator. The $\PZ\PZ\to4\Pell$ events are generated with the dilepton mass $>$1\GeV requirement at the generator level to prevent collinear divergence. All event generators are interfaced with \PYTHIA using the CP5 underlying event tune~\cite{cpxTunes} to describe parton showering and hadronization. The \textsc{NNPDF}3.1 set of parton distribution functions at next-to-next-to-leading order is used~\cite{nnpdf}.
The detector response is simulated using the \GEANTfour~\cite{geant4} package. The effect of additional $\Pp\Pp$ interactions in the same or adjacent bunch crossings (pileup) is incorporated into the simulation by including minimum bias $\Pp\Pp$ interactions generated by \PYTHIA. The simulated samples are reweighted to reproduce the same pileup distribution as measured in data.

Events are collected using nonisolated-lepton triggers that are efficient for both resolved and merged leptons. In the channels with electron-only final states, triggers requiring at least two nonisolated electrons with $\pt > 33$ (25)\GeV are used during the 2016 and 2017 (2018) data-taking periods, with the offline requirement of $\pt > 35$ (28)\GeV to ensure fully efficient triggering. For channels containing at least one muon, single-muon triggers requiring a nonisolated muon having $\pt > 50\GeV$ were used in 2016--2018, with the offline requirement on the leading muon \pt to exceed 52\GeV to ensure efficient triggering.

Electrons with $\pt > 20\GeV$ and $\abs{\eta} < 2.5$ are considered, except the ones falling into the EB-EE gap region at $1.44 < \abs{\eta} < 1.57$.
Electrons are required to pass a modified version of the ID criteria optimized for high energy~\cite{CMS:2020uim} and isolation based on the scalar \pt sum of tracks and calorimeter energy deposits within $\DR\equiv\sqrt{\smash[b]{(\Delta\eta)^{2}+(\Delta\phi)^{2}}} < 0.3$ around the electron. The track and ECAL energy deposits compatible with an overlapping electron are excluded from the isolation calculation~\cite{2017563}. 
Overlapping energy deposits in the ECAL can alter the shower shape and affect the track-cluster geometrical compatibility by shifting the energy-weighted position (\vecClus) of the cluster.
Therefore, new variables are introduced using the extrapolated position of the two electron trajectories on the ECAL surface (\vecEi{1} and \vecEi{2}, ordered by \pt). The shower shape requirement is replaced with a selection based on the angular separation (${\Delta}u$) between \vecClus and its closest approach to the line (\vecEE) connecting \vecEi{1} and \vecEi{2}, which must be $\abs{{\Delta}u} < 0.004$.
The track of the overlapping electron must satisfy $\pt > 10\GeV$ and form a common vertex with the other electron with a vertex fit $\chi^2$ probability $>$1\%.

The clustering algorithm may fail to resolve energy deposits and merge clusters when electrons are highly collimated within $\DR < 0.1$, referred to as \PeME.
In addition to the criteria used to identify resolved electrons, we employ a dedicated $\PeME$ ID based on two boosted decision tree classifiers~\cite{mergedEle}, considering either the case with two tracks or a single track. The two-track model utilizes the separation between \vecClus and \vecEi{1} along \vecEE. The recoil direction, defined by the angle between \vecEE and the azimuthal $\phi$ axis, is also used.
For the single-track model, the ratio of SC energy to the electron track momentum ($E/p$) is used. The two-track (single-track) model is optimized to retain 10\% background efficiency for prompt single electrons, resulting in about 80\,(70)\% \PeME efficiency. The energy of the two-track \PeME is estimated using the energy deposited in the union of the two matrices of $5\times5$ ECAL crystals around the two tracks (\ufxf) since the SC may not contain all energy deposits from the electron pair, as clusters are optimized to be built along the direction perpendicular to the magnetic field.
The \ufxf cluster (SC) energy and the vector to its energy-weighted position from the interaction point are used to estimate the four-momentum of a two-track (single-track) \PeME with the zero-mass hypothesis. 
Electrons in the EB region are considered for the \PeME ID.

Muons must satisfy $\pt > 20\GeV$ and $\abs{\eta} < 2.4$. The \pt sum of tracks within an isolation cone of $\DR < 0.3$ around the muon is required to be $<$10\% of the muon \pt. The \pt of the neighboring muon is subtracted from the isolation sum when the two cones overlap. The simultaneous track reconstruction using both the silicon tracker and muon detectors (``global muon'') may fail if two muons are collimated with $\DR < 0.1$. The resulting efficiency loss is mitigated by using tracks built only with the silicon tracker that are matched to a local trajectory in the muon detectors (``tracker muon''). At least one global muon is required for each dimuon pair. The high-\pt (tracker) muon ID criteria~\cite{CMShighPtMuon} are used to select global (tracker) muons.

For highly collinear dimuons ($\DR < 0.01$), it is even possible for silicon tracker hits to be shared. The trajectory ``cleaner'' merges track candidates with a sizable fraction of shared hits and keeps only the muon with a larger number of hits~\cite{CMS:2014pgm}, resulting in a 40\,(90)\% dimuon reconstruction efficiency at $\DR = 0.001$ (0.01). The removal of one of the two muons (a missing muon, \PmMM) shows no significant effect on the \pt of the remaining muon. Therefore, the presence of \PmMM is inferred from \ptvecmiss, which should point close to the remaining muon in the $\phi$ direction.

The analysis uses several signal regions (SRs) with different lepton flavors and objects. Each SR is defined according to the number of resolved and merged leptons, in addition to a \ptmiss requirement. 
Each SR requires an explicit number of identified leptons to prevent the overlap, as shown in Table~\ref{tab:analysisStrategy}.
The resolved channels comprise SRs with four well-identified leptons, while merged channels include at least one \PeME or \PmMM, where the latter is selected using \ptmiss.
To select $\PX\to\PY\PY$ and $\PX\to\PZ\PY$ signal events simultaneously, no dilepton-mass requirement is applied, except in the resolved channels ($\Mtl > 1\GeV$) to match the $\PZ\PZ\to4\Pell$ MC simulation used for estimating irreducible prompt-lepton backgrounds.

\begin{table}[htb]
\centering
\topcaption{The SRs, where each region is defined by the listed reconstructed leptons in an event. The \Pe and \PGm represent resolved leptons, while \PeME and \PmMM represent the merged electron and missing muon, respectively.\label{tab:analysisStrategy}}
\renewcommand{\arraystretch}{1.1}

\begin{scotch}{lccc}
     Resolved & 4\Pe & 2\Pe~2\PGm & 4\PGm\\
     \hline
     \multirow{3}{*}{Merged} & 
       \multirow{3}{*}{
         \begin{tabular}{@{}c@{}} 
           2\Pe~\PeME \\ 2\PeME 
         \end{tabular}
       } & \PeME~2\PGm & \\
     & & \PeME~\PGm~\PmMM & 3\PGm~\PmMM\\
     & & 2\Pe~\PGm~\PmMM &
\end{scotch}
\end{table}

In the resolved channels, electrons with $\DR < 0.4$ are paired according to the two-track model; otherwise the pair with the closest \DR is chosen to resolve the ambiguity. Muons are always paired according to the closest \DR.

In the merged channels, the invariant mass of all visible leptons must be $>$50\GeV. 
For the channels with a missing muon, \ptmiss must be $>$50\GeV and the $\abs{\Delta\phi}$ between \ptvecmiss and the closest muon must be $<$0.3. Because \ptvecmiss is used as a proxy for the \PmMM, the ratio of the vector sum of visible leptons to \ptmiss (\ptratio) is expected to be close to unity. We require the ratio to be in the range 0.5--1.5, corresponding to about 70\% \PmMM efficiency. The transverse mass ($\mT = \sqrt{\smash[b]{E^2-p_{z}^2}}$) 
must be $>$250\GeV, where $E$ and $p_{\mathrm{z}}$ are the energy and the $z$ component of the momentum of the system formed by the visible leptons and \ptmiss.
For the $3\PGm\PmMM$ channel, a collimated dimuon pair with $\DR < 0.3$ is required. 
The dimuon pair with the smallest \DR is chosen and the remaining muon is paired with \ptmiss.

The estimation of the background considers three major categories: the resolved channels, the merged channels with \PeME but without \ptmiss, and the merged channels with \ptmiss. The irreducible prompt-lepton background from $\PZ\PZ\to4\Pell$ events in the resolved channels is estimated using MC simulation. Other nonprompt-lepton backgrounds, caused by the misidentification of either \Pell, \PeME, or \PmMM, where \Pell represents either a resolved electron or muon, are estimated using misidentification factors derived from the ratio of event yields in dedicated control regions (CRs) shown in Table~\ref{tab:fakeFactorCR}. The background predictions in all categories are cross-checked using validation regions in data.

\begin{table*}[htb]
\centering
\topcaption{A summary of the CRs used to derive the misidentification factors for different lepton classes. The \PZ represents an SM \PZ boson candidate, formed by two opposite-sign same-flavor resolved leptons passing the ID criteria with an invariant mass of $91\pm7\GeV$, and $\Pell^{\prime}$ denotes a resolved lepton with a different flavor (\Pe or \PGm) than \Pell.\label{tab:fakeFactorCR}}
\renewcommand{\arraystretch}{1.1}
\begin{scotch}{llcc}
Class & Subcategory & Numerator & Denominator\\
\hline
\multirow{2}{*}{$\Pell$} & Nonprompt ($\DR > 0.3$) & $\PZ + \PellPass$ & $\PZ + \PellFail$\\
& Nonprompt ($\DR < 0.3$) & $2\PellPass + \PellPass^{\prime}$ & $2\PellFail + \PellPass^{\prime}$\\[\cmsTabSkip]

\multirow{2}{*}{$\PeME$} & Prompt background & $\PeME + \PeMEFail$ & $2\PeMEFail$\\
& Nonprompt background & $\PZ + \PeME$ & $\PZ + \PeMEFail$\\[\cmsTabSkip]

$\PmMM$ & & $\ptratio \in (0.5, 1.5)$ & $\ptratio \notin (0.5, 1.5)$\\
\end{scotch}
\end{table*}

The nonprompt-lepton background in the resolved channels originates from Drell-Yan (DY) and $\PQt\PAQt$ events with additional misidentified leptons, which can be categorized into isolated ($\DR > 0.3$) and nonisolated ($\DR < 0.3$) components, based on the $\DR$ between the two closest same-flavor lepton candidates.
For the isolated case, the misidentification factor is measured using the region with two opposite-sign same-flavor leptons satisfying the ID criteria and an additional lepton candidate~\cite{CMShiggs2017}. The two well-identified leptons form a \PZ boson candidate within an invariant mass window within $91\pm7\GeV$ (``\PZ tagging'') that is wide enough to account for the lepton momentum scale and resolution effect. 
Misidentification factors are derived using the ratio of the number of lepton candidates passing the signal ID criteria (\PellPass) to those passing only looser ID criteria (\PellFail). 
The isolated nonprompt-lepton backgrounds are estimated by applying the misidentification factor to the number of four-lepton events with at least two of type \PellPass and at least one \PellFail, where each paired dilepton has $\DR > 0.3$.

For nonisolated lepton pairs, the misidentification factor is measured using data enriched in $\PW\to\Pell\PGn$ and $\Pell+\mathrm{jets}$ $\PQt\PAQt$ events with a misidentified dilepton pair. The dilepton pair must have $\DR < 0.3$, and an additional well-identified lepton of a different flavor is required. A $\ptmiss > 50\GeV$ selection is applied to suppress DY events and ensure that the dilepton candidates are not prompt leptons. The dilepton misidentification factor is calculated as the number of \PellPass pairs divided by the number of \PellFail pairs. The nonisolated nonprompt-lepton backgrounds are estimated by applying the dilepton misidentification factor to the number of $2\PellPass2\PellFail$ events with $\DR < 0.3$ for each pair.

In the merged channels containing only \PeME, the misidentified \PeME candidates can originate from both nonprompt and prompt single-electron contributions. The \PeME object misidentification factor from prompt electrons is measured using opposite-sign dielectron events with $50 < m_{\Pe\Pe} < 200\GeV$. The prompt misidentification factor is defined as the number of events with one passing \PeME and one failing \PeME candidate (\PeMEFail) divided by the number of 2\PeMEFail events. The background in the 2\PeME SR is dominated by the prompt contribution and estimated by applying the prompt-\PeME misidentification factor to the number of events with $m(\PeME,\PeMEFail) > 200\GeV$.
The misidentification factor from the nonprompt-lepton contribution is measured using the CR with an \PeME candidate and a tagged \PZ boson in data. The prompt-electron contamination from \ttbar, $\PW\PZ\to3\Pell\PGn$, and partially reconstructed $\PZ\PZ\to4\Pell$ is subtracted using MC simulation. The nonprompt-\PeME misidentification factor is defined as the ratio of the number of events with \PeME to \PeMEFail.

For the $2\Pell\PeME$ SRs, both prompt- and nonprompt-lepton backgrounds are nonnegligible. The prompt-electron contributions are derived from multiplying the prompt misidentification factor by the number of simulated \ttbar, $\PW\PZ\to3\Pell\PGn$, and $\PZ\PZ\to4\Pell$ events with $2\PellPass\PeMEFail$. Events with $m(\Pe,\PeMEFail)$ within $91\pm7\GeV$ are also considered in the prompt contribution. The nonprompt-lepton background is estimated by multiplying the nonprompt-\PeME misidentification factor by the number of events with $2\PellPass\PeMEFail$, after subtracting the number of prompt-lepton background events using MC simulation.

The background for the \PmMM SRs, including $\PeME\PGm\PmMM$, is estimated using the ratio of the number of events passing the $0.5 < \ptratio < 1.5$ selection to the number failing the selection. The ratio is derived using events with $\abs{\Delta\phi}$ between \ptvecmiss and the nearby muon $>$0.6 and is validated using the CR with $0.3 < $\abs{\Delta\phi}$ < 0.6$. Then, the ratio is applied to the number of events that fail the \ptratio selection but satisfy $\abs{\Delta\phi} < 0.3$. Events with $\mT > 800$ (1000)\GeV with an additional dimuon (dielectron) pair are counted in the same \mT bin.

The uncertainty is dominated by the low number of events observed in the SRs. The largest systematic uncertainty in the merged channels is the \PeME ID efficiency, which contributes an uncertainty of about 6\,(30)\% to the background yield for the two-track (single-track) model. The \PeME efficiency uncertainty is derived using Lorentz-boosted \PJGy mesons decaying to \EE and $\PZ\to\PGmp\PGmm\PGg$ events where the photon converts into an \EE pair.
The electron trigger efficiency uncertainty for \PeME is approximately 3\%.
The \ufxf cluster energy scale and resolution uncertainties are estimated using an invariant mass of the cluster and a track from $\PBpm\to\PJGy(\to\EE)\PKpm$ events~\cite{mergedEle}. The \PeME energy uncertainty is 0.3\% for the scale and 2\% for the resolution.

The prompt-\PeME misidentification factor is affected by the resolution of the $E/p$ variable in the single-track model, resulting in 10\,(20)\% uncertainty at $\pt = 200$ (1000)\GeV. The nonprompt-\PeME misidentification factor is driven by the two-track model and shows no \pt dependence, with about 20\% uncertainty.
The \PmMM background estimation uncertainty is about 15\%. The shared silicon tracker hits of the \PmMM contribute an uncertainty of 4\,(15)\% to the muon momentum scale (resolution).

The uncertainties in the resolved-lepton reconstruction, ID, and trigger efficiencies, in addition to energy scale and resolution, are determined using $\PZ\to\Pell^{+}\Pell^{-}$ events \cite{CMS:2020uim,CMShighPtMuon}. A deviation of about 10\% is observed for the irreducible prompt-lepton background in the $\Mfl < 200\GeV$ CR, and a corresponding uncertainty is assigned to the $\PZ\PZ\to4\Pell$ MC sample normalization. The isolated-lepton (nonisolated-dilepton) misidentification factor decreases exponentially as a function of the lepton (dilepton) \pt. Its uncertainty is about 5--10\%, mainly arising from the low event counts in the CRs. The uncertainty in the integrated luminosity measurement for each data-taking period is included~\cite{LUM-17-003,LUM-17-004,LUM-18-002}. Lastly, the uncertainty on pileup reweighting is estimated by recalculating weights with the total inelastic cross section shifted by $\pm4.6\%$~\cite{CMS:2018mlc}. Systematic uncertainties are incorporated in the likelihood fit of the \Mfl and \mT distributions, performed using the modified frequentist method~\cite{junk1999435} and the CMS statistical analysis tool \textsc{Combine}~\cite{CMS:2024onh}, based on the \textsc{RooFit}~\cite{Verkerke:2003ir} and \textsc{RooStats}~\cite{Moneta:2010pm} frameworks.

\begin{figure*}[ht]
  \centering
    \includegraphics[width=0.99\textwidth]{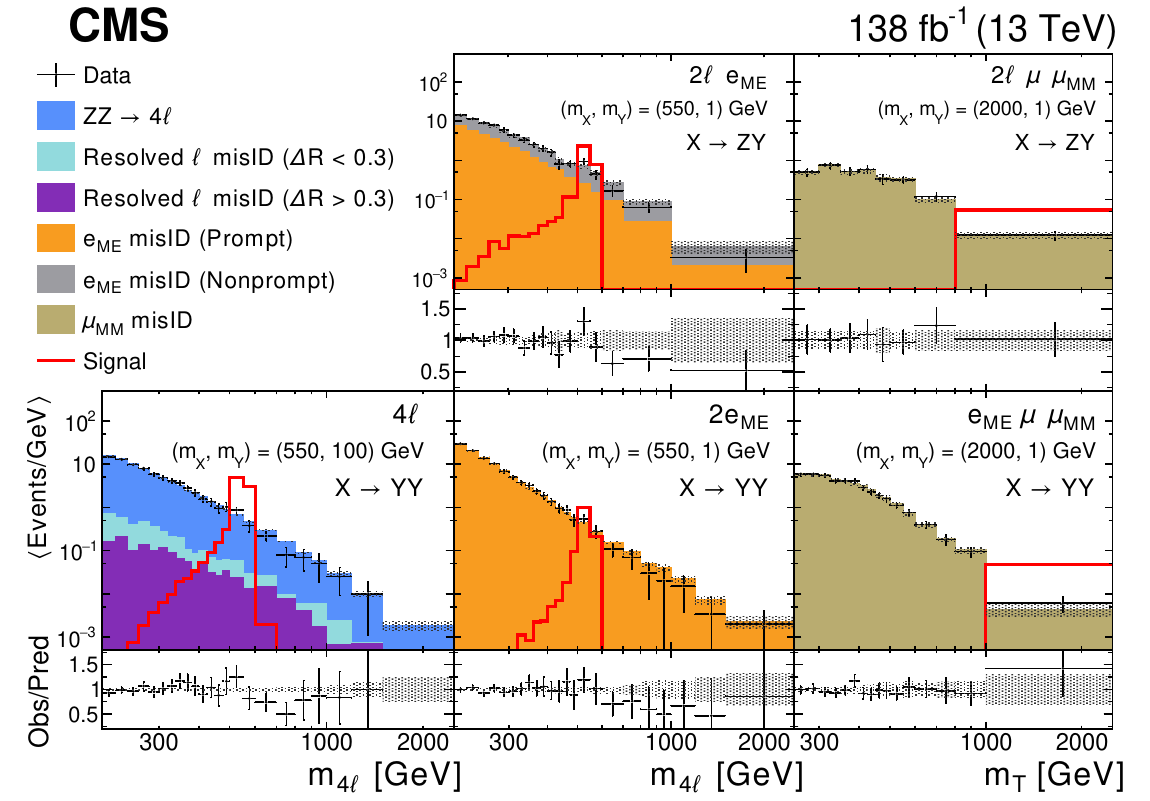}
    \caption{The observed and predicted \Mfl (left and center) and \mT (right) distributions after the likelihood fit under the background-only hypothesis. The lower panel shows the ratio of the number of observed events to the predicted background. The shaded band represents the postfit background uncertainty, incorporating both systematic and statistical sources. The red histogram shows the expected signal distribution for a given $(\MX,~\MY)$ and decay-mode hypothesis, normalized to $\sigma(\Pp\Pp\to\PX)\mathcal{B}(\PX\to4\Pell) = 10\unit{fb}$.}
    \label{fig:signalRegions}
\end{figure*}

The \Mfl (\mT) distributions for the SRs without (with) \ptmiss after the likelihood fit under the background-only hypothesis are shown in Fig.~\ref{fig:signalRegions}. No significant excess of data over the SM background predictions is observed. Upper limits on the cross section times branching fraction $\sigma(\Pp\Pp\to\PX)\mathcal{B}(\PX\to\PY\PY\to4\Pell)$ and $\sigma(\Pp\Pp\to\PX)\mathcal{B}(\PX\to\PZ\PY\to4\Pell)$ are set.

\begin{figure*}[ht]
  \centering
    \includegraphics[width=0.99\textwidth]{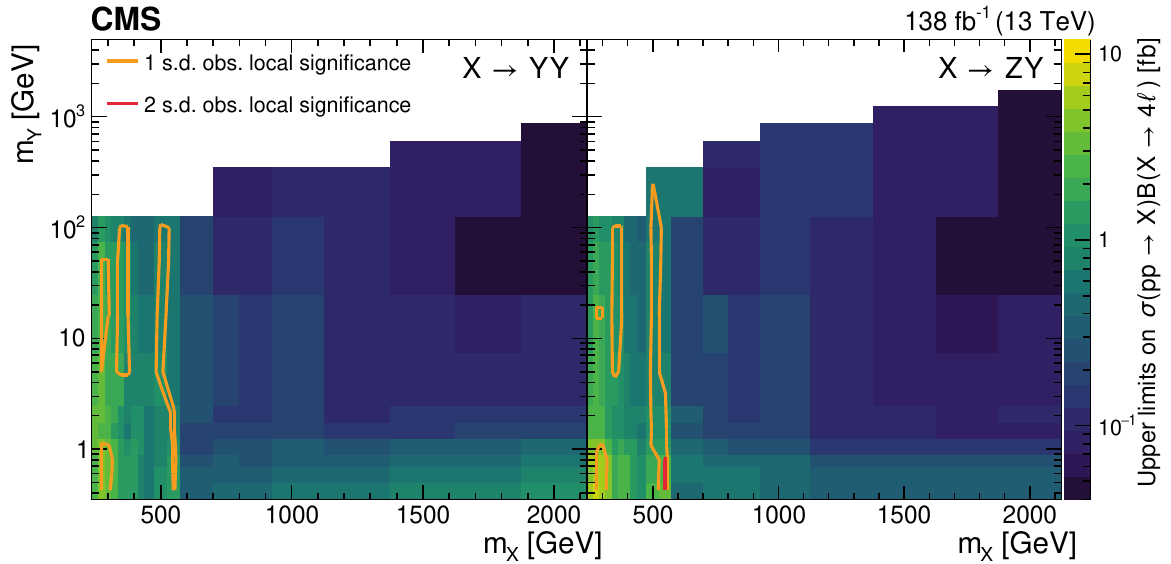}
    \caption{The observed limit on the product of the cross section and branching fraction for (left) $\PX\to\PY\PY\to4\Pell$ and (right) $\PX\to\PZ\PY\to4\Pell$ signal hypotheses. The orange (red) solid line represents the 1 (2) \SD observed local significance contour. The white area indicates a kinematically forbidden region.}
    \label{fig:limit2D_PRL}
\end{figure*}

Figure~\ref{fig:limit2D_PRL} shows the observed 95\% confidence level upper limits as functions of \MX and \MY for each decay mode. 
The local significance of rejecting a background-only hypothesis over the signal+background hypothesis is tested for various signal mass points.
The highest local significance of 2.1 (1.9) standard deviations (\SD) is observed at $\MX = 550\GeV$ and $\MY = 0.8\GeV$ for the $\PX\to\PZ\PY$ ($\PX\to\PY\PY$) decay mode, originating from the $2\PGm\PeME$ SR. A global significance is also estimated, including the look-elsewhere effect~\cite{LEE} by counting the number of background-only simulation trials with significance greater than the observed significance in data. The estimated global significance is 1.2 (0.9) \SD within the range $250 < \MX < 2000\GeV$ for the $\PX\to\PZ\PY$ ($\PX\to\PY\PY$) channel.
The upper limits for exclusive $\PY\to\Pe\Pe$ and $\PY\to\PGm\PGm$ decay modes are available in \appendixname~\ref{app:endMat}.
Tabulated results of signal efficiencies, four-lepton and \mT distributions, and upper limits are available in the \textsc{HEPData} record~\cite{hepdata}.

In summary, a model-independent search for a resonance heavier than 250\GeV decaying into four-lepton final states has been presented for the $\PX\to\PY\PY\to4\Pell$ and $\PX\to\PZ\PY\to4\Pell$ channels, where \PX and \PY are hypothetical new bosons with masses \MX and \MY. The search uses proton-proton collision data at $\sqrt{s} = 13\TeV$ collected with the CMS detector at the LHC, corresponding to an integrated luminosity of 138\fbinv. Decays of a light \PY boson with a mass $\lesssim$10\GeV can result in final states with a large Lorentz boost, producing highly collimated dielectron or dimuon signatures.
A dedicated merged-electron identification is developed to enhance signal sensitivity to collimated dielectrons from a low-mass \MY.
Events with collimated dimuons are analyzed using the missing transverse momentum vector as a proxy for the missing muon. The maximum local (global) observed significance is 2.1 (1.2) standard deviations at $\MX = 550\GeV$ and $\MY = 0.8\GeV$ in the $\PX\to\PZ\PY\to4\Pell$ channel. The 95\% confidence level upper limits on the product of cross section and branching fraction to 4\Pell final states are set for $\MX > 250\GeV$ and $\MY > 0.4\GeV$. 
This is the first LHC result probing 4\Pell resonances heavier than the standard model Higgs boson that decay via intermediate bosons with $0.4 < \MY < 15\GeV$.

\begin{acknowledgments}
We congratulate our colleagues in the CERN accelerator departments for the excellent performance of the LHC and thank the technical and administrative staffs at CERN and at other CMS institutes for their contributions to the success of the CMS effort. In addition, we gratefully acknowledge the computing centers and personnel of the Worldwide LHC Computing Grid and other centers for delivering so effectively the computing infrastructure essential to our analyses. Finally, we acknowledge the enduring support for the construction and operation of the LHC, the CMS detector, and the supporting computing infrastructure provided by the following funding agencies: SC (Armenia), BMBWF and FWF (Austria); FNRS and FWO (Belgium); CNPq, CAPES, FAPERJ, FAPERGS, and FAPESP (Brazil); MES and BNSF (Bulgaria); CERN; CAS, MoST, and NSFC (China); MINCIENCIAS (Colombia); MSES and CSF (Croatia); RIF (Cyprus); SENESCYT (Ecuador); ERC PRG and PSG, TARISTU24-TK10 and MoER TK202 (Estonia); Academy of Finland, MEC, and HIP (Finland); CEA and CNRS/IN2P3 (France); SRNSF (Georgia); BMFTR, DFG, and HGF (Germany); GSRI (Greece); MATE and NKFIH (Hungary); DAE and DST (India); IPM (Iran); SFI (Ireland); INFN (Italy); MSIT and NRF (Republic of Korea); MES (Latvia); LMTLT (Lithuania); MOE and UM (Malaysia); BUAP, CINVESTAV, CONACYT, LNS, SEP, and UASLP-FAI (Mexico); MOS (Montenegro); MBIE (New Zealand); PAEC (Pakistan); MSHE, NSC, and NAWA (Poland); FCT (Portugal); MESTD (Serbia); MICIU/AEI and PCTI (Spain); MOSTR (Sri Lanka); Swiss Funding Agencies (Switzerland); MST (Taipei); MHESI (Thailand); TUBITAK and TENMAK (T\"{u}rkiye); NASU (Ukraine); STFC (United Kingdom); DOE and NSF (USA).
\end{acknowledgments}

\bibliography{auto_generated}

\appendix
\section{Upper limits for exclusive \texorpdfstring{$\PY\to\Pe\Pe$}{Y to ee} and \texorpdfstring{$\PY\to\mu\mu$}{Y to mumu} decay modes}\label{app:endMat}
\begin{figure*}[ht]
  \centering
    \includegraphics[width=0.8\textwidth]{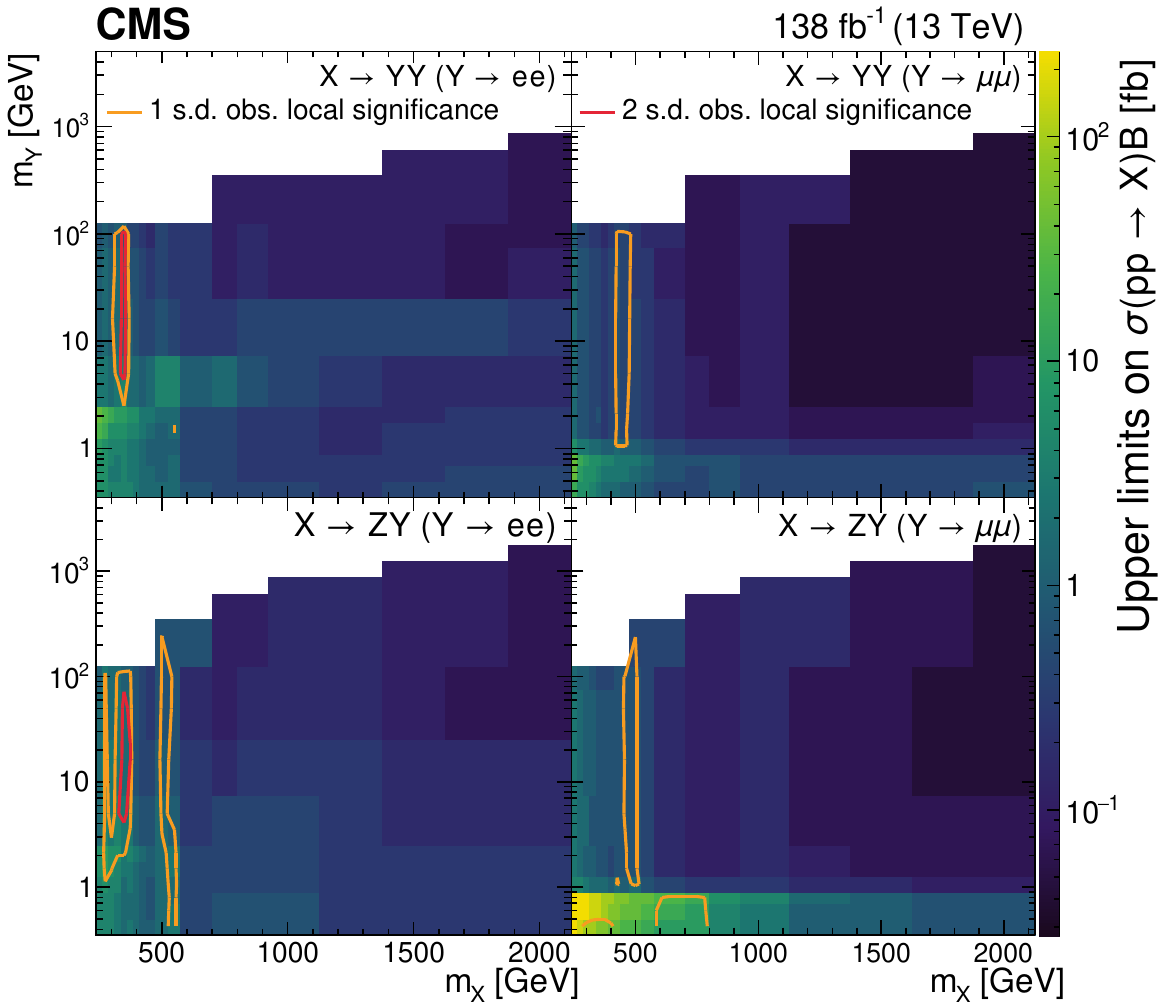}
    \caption{The observed limit on the product of the cross section and branching fraction for (upper left) $\PX\to\PY\PY\to4\Pe$, (upper right) $\PX\to\PY\PY\to4\PGm$, (lower left) $\PX\to\PZ\PY\to2\Pell2\Pe$, and (lower right) $\PX\to\PZ\PY\to2\Pell2\PGm$ signal hypotheses, where the \PZ boson decays into either a dielectron or dimuon. The orange (red) solid line represents the $1$ ($2$) \SD observed local significance contour. The white area indicates a kinematically forbidden region.}
    \label{fig:limit2D_PRL_supp}
\end{figure*}

The lepton universality between $\PY\to\Pe\Pe$ and $\PY\to\PGm\PGm$ may not be guaranteed in some models~\cite{2HDMreview}. The observed upper limits on the cross section times branching fraction with exclusive $\PY\to\Pe\Pe$ and $\PY\to\PGm\PGm$ decays are also examined and shown in Fig.~\ref{fig:limit2D_PRL_supp}. For the $\PX\to\PZ\PY$ hypothesis, the SM \PZ boson is set to decay into either dielectron or dimuon pairs for both $\PY$ decay scenarios.

Figure~\ref{fig:limit2D_PRL_supp} (lower left) shows the local (global) significance of 1.8 (1.2) \SD at $(\MX,~\MY) = (550,~0.8)\GeV$ visible exclusively in the $\PX\to\PZ\PY\to2\Pell2\Pe$ decay.
Other local (global) significances are observed at $(\MX,~\MY) = (350,~50)\GeV$ with 2.7 (1.5) \SD for $\PX\to\PY\PY\to4\Pe$ and 2.1 (0.7) \SD for $\PX\to\PZ\PY\to2\Pell2\Pe$ decays. The $\PY\to\PGm\PGm$ channel provides a more stringent limit than $\PY\to\Pe\Pe$ at $\MY > 1\GeV$, because resolved muons have higher reconstruction and ID efficiencies than electrons. In contrast, the $\PY\to\Pe\Pe$ channel yields a more stringent limit than $\PY\to\PGm\PGm$ at $\MY < 1\GeV$, as the \PeME selection efficiency is higher than \PmMM.

\section{Dilepton mass distribution of the merged channels}

\begin{figure*}[ht]
  \centering
    \includegraphics[width=0.9\textwidth]{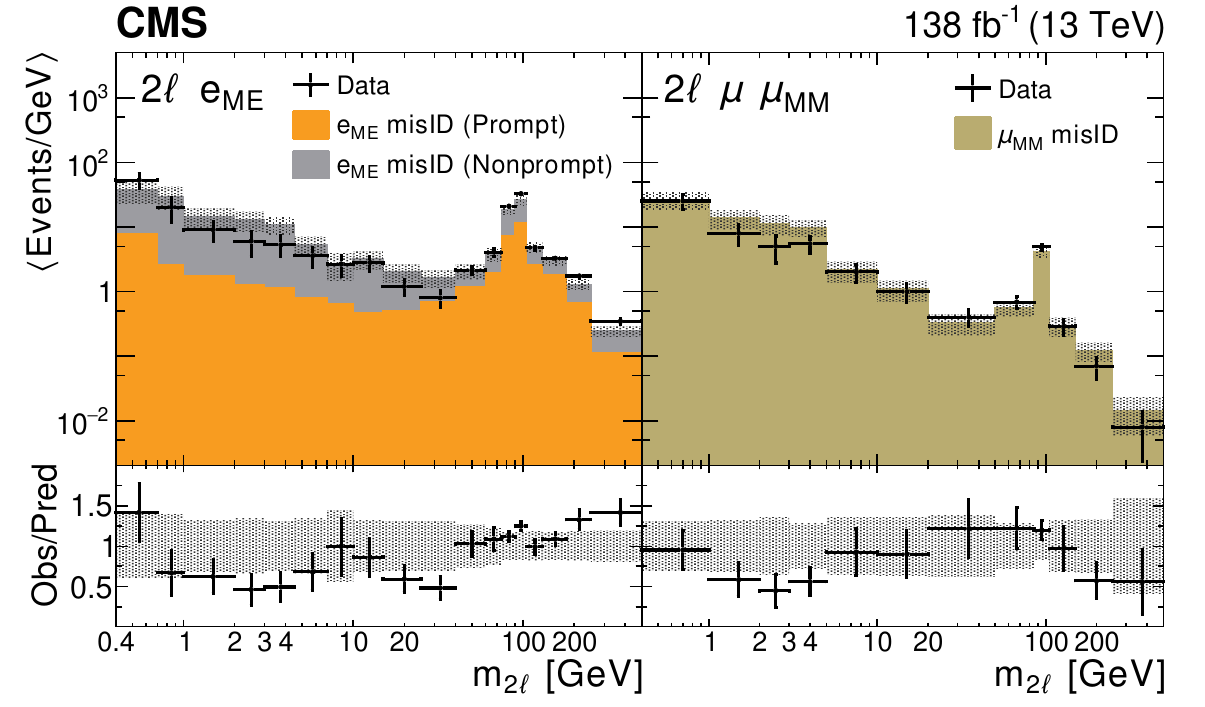}
    \caption{The observed and predicted \Mtl distribution in the (left) $2\Pell\PeME$ and (right) $2\Pell\PGm\PmMM$ channels, before the likelihood fit. The shaded band represents the combined systematic and statistical uncertainties, summed in quadrature. The lower panel shows the ratio of the number of observed events to the predicted background.}
    \label{fig:dileptonMass}
\end{figure*}

The invariant mass of the merged dilepton is difficult to reconstruct reliably; hence, \MY is not used in the likelihood fit. However, in the $2\Pell\PeME$ and $2\Pell\PGm\PmMM$ SRs, it is possible to reconstruct the invariant mass of the accompanying resolved lepton pair (\Mtl). The \Mtl distributions before the likelihood fit are shown in Fig.~\ref{fig:dileptonMass}. For the $\PX\to\PZ\PY$ signal hypothesis, \Mtl peaks near the \PZ boson mass, whereas \Mtl is expected to be $<$10\GeV for $\PX\to\PY\PY$.

\cleardoublepage \section{The CMS Collaboration \label{app:collab}}\begin{sloppypar}\hyphenpenalty=5000\widowpenalty=500\clubpenalty=5000\cmsinstitute{Yerevan Physics Institute, Yerevan, Armenia}
{\tolerance=6000
A.~Hayrapetyan, V.~Makarenko\cmsorcid{0000-0002-8406-8605}, A.~Tumasyan\cmsAuthorMark{1}\cmsorcid{0009-0000-0684-6742}
\par}
\cmsinstitute{Institut f\"{u}r Hochenergiephysik, Vienna, Austria}
{\tolerance=6000
W.~Adam\cmsorcid{0000-0001-9099-4341}, L.~Benato\cmsorcid{0000-0001-5135-7489}, T.~Bergauer\cmsorcid{0000-0002-5786-0293}, M.~Dragicevic\cmsorcid{0000-0003-1967-6783}, P.S.~Hussain\cmsorcid{0000-0002-4825-5278}, M.~Jeitler\cmsAuthorMark{2}\cmsorcid{0000-0002-5141-9560}, N.~Krammer\cmsorcid{0000-0002-0548-0985}, A.~Li\cmsorcid{0000-0002-4547-116X}, D.~Liko\cmsorcid{0000-0002-3380-473X}, M.~Matthewman, J.~Schieck\cmsAuthorMark{2}\cmsorcid{0000-0002-1058-8093}, R.~Sch\"{o}fbeck\cmsAuthorMark{2}\cmsorcid{0000-0002-2332-8784}, M.~Shooshtari\cmsorcid{0009-0004-8882-4887}, M.~Sonawane\cmsorcid{0000-0003-0510-7010}, W.~Waltenberger\cmsorcid{0000-0002-6215-7228}, C.-E.~Wulz\cmsAuthorMark{2}\cmsorcid{0000-0001-9226-5812}
% ...
\par}
\vskip\cmsinstskip
\dag:~Deceased\\
$^{1}$Also at Yerevan State University, Yerevan, Armenia\\
$^{2}$Also at TU Wien, Vienna, Austria\\
% ...
\end{sloppypar}
\end{document}